# The Electromagnetic Barrel Calorimeter for the GlueX Experiment

M. Barbi

*Department of Physics, University of Regina*
*3737 Wascana Parkway, Regina, SK, S4S 0A2, Canada*

**Abstract.** The electromagnetic barrel calorimeter is one of the main components of the planned GlueX experiment. It will consist of 48 modules made of consecutive layers of 4 m long lead sheet and fast green scintillator fibers for an overall number of approximately 3000 readout channels with silicon-photomultiplier-based photo-sensors for light collection. The calorimeter is expected to achieve energy and time resolution better than $5\%/\sqrt{E} \oplus 2\%$ and 200 ps, respectively.

In this contribution we present an overview of the calorimeter design and some preliminary studies of its performance using Monte Carlo simulations and beam test measurements.

## INTRODUCTION

The Standard Model (SM) is the most successful theory to date to explain the fundamental building blocks which make all the existing matter in the Universe and the forces through which they interact. Within the SM, the Quantum Chromodynamics theory (QCD) describes the strong interaction that bounds quarks and gluons together inside the nucleons. QCD has succeeded in explaining the quark-quark interactions at short distance scale, where the strong coupling constant $\alpha_s$ is small and perturbative QCD calculations apply. However, at large distance scales, $\alpha_s$ is large, perturbative calculations no longer apply and the complicated QCD Lagrangians are difficult to solve. However, it is at these scales that the QCD confinement mechanism which prevents quarks from escaping the strong potential generated by the gluonic fields plays its role in the nucleon structure. One way of working around the non-perturbative regime is by using Lattice QCD calculations [1]. Another way is to rely on physics models in an attempt to approximately reproduce the quark-gluon dynamics at large $\alpha_s$ values. In either cases, the difficult in solving the QCD Lagrangian in the non-perturbative regime leads to one of the more challenging and outstanding problems still to be solved in QCD, that of unfolding the real nature of the QCD confinement mechanism.

The Flux-Tube Model [2] is one of the most accepted models to describe the quark-gluon dynamics at large distance scales, in which quarks are bounded to each other through gluonic flux tubes. Regular quark bound states happen when the flux tubes are in their ground state. However, when in excited mode, the gluonic flux tubes can add degrees of freedom to the quark system, consequently leading to final states where gluons also contribute to the total quantum numbers. These states, where not only quarks but also gluons effectively participate in the full dynamics, are called hybrid states. Quark-gluon bound states are allowed by QCD, though they are not permitted in the regular Quark Model.

The GlueX experiment (Figure 1) is a project of a detector being developed dedicated to map the hybrid mesons states using the CEBAF accelerator facilities at the Thomas Jefferson Laboratory (JLAB). The data will be collected from fixed-target photoproduction in γp collisions using 9 GeV coherent Bremsthalung photons generated from a primary 12 GeV electron beam source. The final states will be fully established using Partial Wave Analysis (PWA) techniques. To be effective and unbiased, PWA requires GlueX to have a very large acceptance to fully contain the particles produced in γp reactions, and to have high energy, momentum and position resolution to allow for reconstruction of the four-momentum of individual particles to high accuracy.

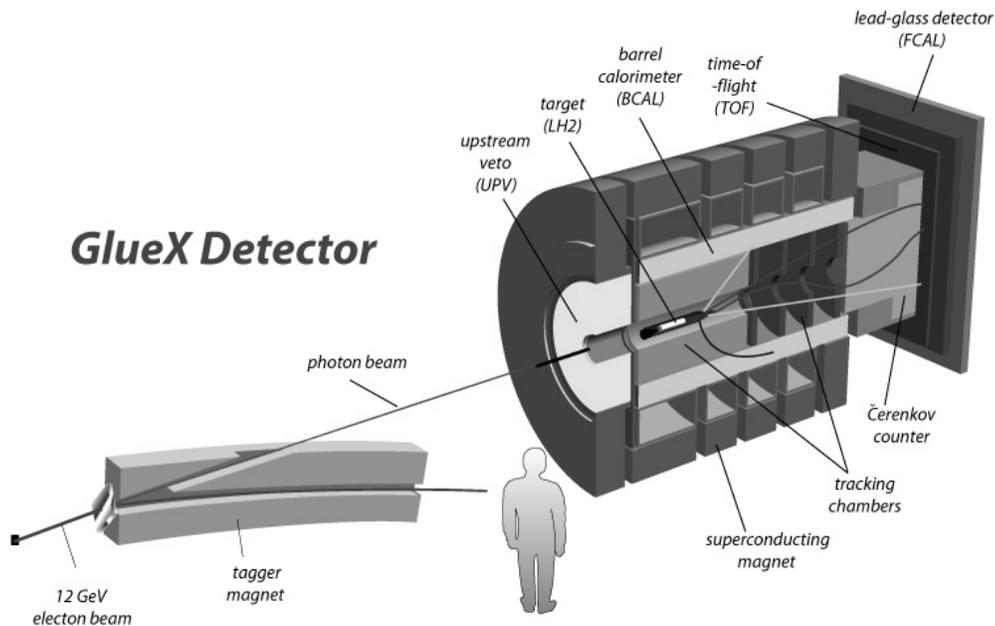

**FIGURE 1.** The GlueX detector.

The electromagnetic barrel calorimeter (BCAL) is one of main components of the GlueX detector. Its primary goal is to detect photons from decays of $\pi^0$ and η particles. The calorimeter is made of consecutive layers of 4 m long lead (Pb) sheet and scintillator fibers (SciFi) acting as passive and active material respectively, and will consist of 48 modules as depicted in Figure 2. The SciFi's are laid into grooves machined longitudinally in the lead sheets relative to the incoming photon beam. The tracking chambers and superconducting solenoid constrain the inner and outer radius of the BCAL to 65 cm and 90 cm respectively, corresponding to an overall acceptance of $12° < \theta < 135°$.

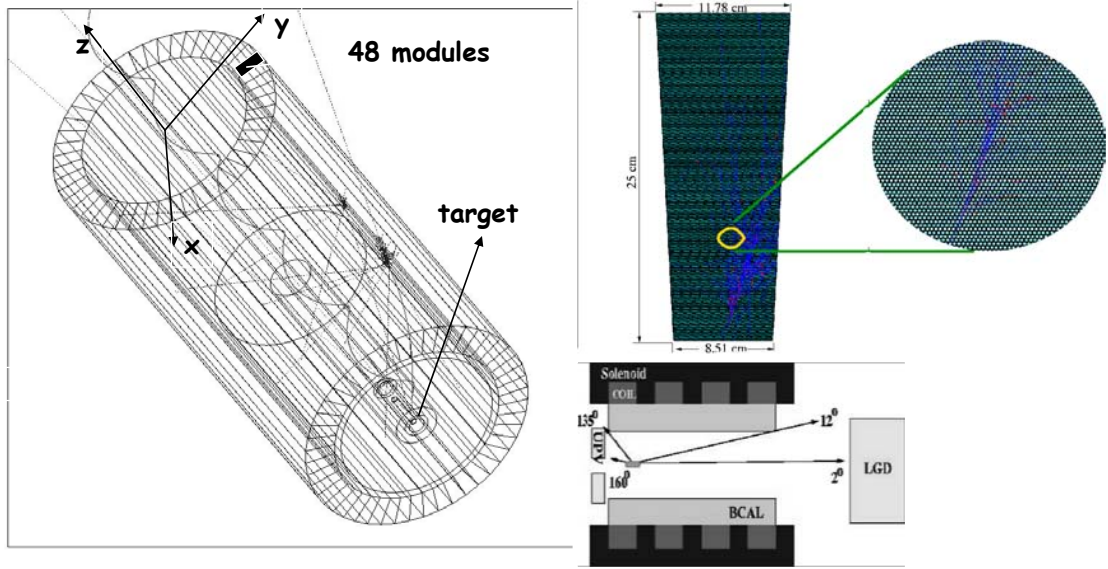

**FIGURE 2.** An overview of the full calorimeter is shown on the left with a simulation of a γp reaction using GEANT-3, while the top right figure is a GEANT-3 simulation of the cross-section of one of the calorimeter modules. The bottom right sketches the calorimeter acceptance region.

The GlueX-BCAL is a non-compensating sampling calorimeter with a Pb:SciFi:glue ratio of 37:49:14. A section of a module is represented in Figure 3 showing the geometrical dimensions of the lead-scintillator layers including the gluing material used to fix the SciFi's into the grooves of the lead sheets.

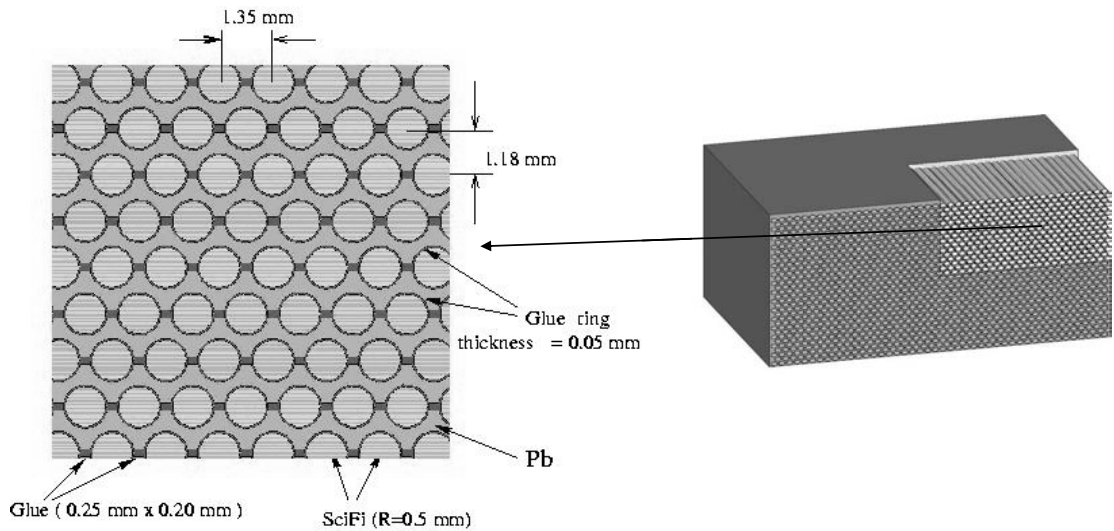

**FIGURE 3.** The inner geometrical structure of the calorimeter.

The BCF-20 double-clad fast green scintillator fibers from BICRON with emission peak at 492 nm, 1 mm diameter, attenuation length greater than 300 cm, decay time of ~ 3 ns and 8000 photons/MeV were chosen as the best candidate to date for the active material of the BCAL. A total of ~3500 Km of fiber equivalent to a density of 63 fibers/cm$^2$ will be needed for the full construction of the calorimeter.

## The GlueX-BCAL Requirements

The designed BCAL energy, position and time resolution are given below:

$$\frac{\sigma(E)}{E} = \frac{5\%}{\sqrt{E}} \oplus 1\% \ ; \qquad \sigma(x) = 1 \ cm \ ; \qquad \sigma(t) = 200 \ ps. \qquad (1)$$

The geometrical and material composition of the calorimeter ensure a thickness of ~15.6 $X_0$ radiation lengths needed to fully contain electromagnetic showers produced by incident particles with energy up to 3 GeV. The resulting Molière radius $R_M \approx 4 \ cm$ requires fine readout segmentation area of at most 4 cm$^2$ to accomplish good shower separation and therefore fulfill the requirements of excellent particle identification and high energy resolution. A possible readout scheme will have the most inner layers of each BCAL module, corresponding to the first section of 10 cm in depth where most of the energy of the income particles is deposited, segmented into 20 readout channels of ~2x2 cm$^2$ area each, while the outer layers corresponding to the second section of 14 cm in depth will have 12 readout channels with area of ~3.5x4 cm$^2$ each, for a total of 64 readout channels per module, 32 for each BCAL side, leading to the overall number of 3072 readout channels for the whole calorimeter.

The GlueX-BCAL will be inserted and operated in a ~2 T magnetic field required for the GlueX experiment. This high magnetic field is a constraint to the choice of the photo-sensors for the calorimeter. Under this condition, regular vacuum PMT's are very sensitive to the orientation of the magnetic field. On the other hand, preliminary studies [4] have shown that silicon photomultipliers (SiPM's) are suitable for the requirements of the calorimeter readout operation, i.e., up to ~1 MHz counting rate, 20 MeV – 3 GeV dynamic range, and high energy and time resolution. These devices are also insensitive to magnetic fields up to at least ~4 T [5], and can deliver high resolution single photo-electron detection at low noise level.

A detailed description of SiPM's can be found elsewhere [6]. Some of their main properties are: high gain (10$^6$-10$^7$ times); low bias voltage normally in the range of 20-60 V; high dynamic range; fast recovery time at the order of few tenths of nanosecond; short rise time at the order of 1-2 ns. Also, currently available SiPM's have higher sensitivity in the green region of the light spectrum. This is most desirable since preliminary studies[1] carried out by the GlueX collaboration have shown that even using long blue scintillator fibers, the light that survives to the ends of the BCAL is in the green region of the spectrum. It has been therefore decided that a better approach would be to use green scintillator fibers coupled to green sensitive photo-sensors.

---

[1] To be submitted for publication.

# Preliminary Studies

Several Monte Carlo simulations have been carried out in order to optimize the calorimeter response and test its potential to reconstruct electromagnetic showers in GlueX. Figure 4 is a GEANT-3 simulation of the process $\gamma p \to \eta \pi^0 p \to \gamma\gamma\gamma\gamma p$, where 9 GeV photons collide with fixed-target protons. The first three distributions show the ratio between the total energy $E_{\gamma,p}$ of the final state and the energy deposited in the lead ($E_{dep}^{Lead}$), SciFi ($E_{dep}^{SciFi}$) and glue ($E_{dep}^{Glue}$) respectively, while the fourth distribution shows the fraction of energy $E^{Leak}$ leaking from the calorimeter. 97% of the total energy produced in this reaction is fully contained in the calorimeter.

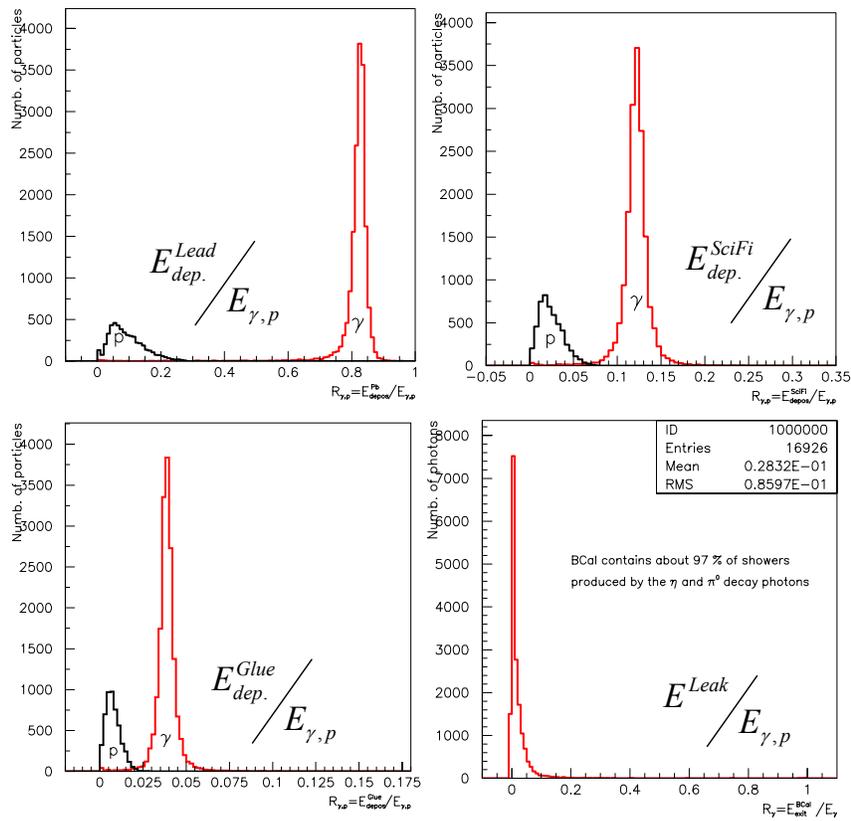

**FIGURE 4.** Distribution of the energy fraction deposited in the lead, scintillator and glue material. The fraction of energy not captured by the calorimeter is also shown in the bottom right distribution.

The reconstructed invariant mass of $\pi^0$ particles simulated with energy of 2 GeV using the proposed BCAL readout scheme as discussed in the previous section implemented into GEANT-3 is shown in Figure 5. The $\pi^0$'s enter the BCAL at random angles. There is no electronics simulation included and the reconstruction software is a very preliminary version which is still being optimized for events with more than one electromagnetic shower. The large background at lower invariant masses is due to photons being associated to wrong showers. This background is expected to significantly improve with the use of optimized reconstruction software.

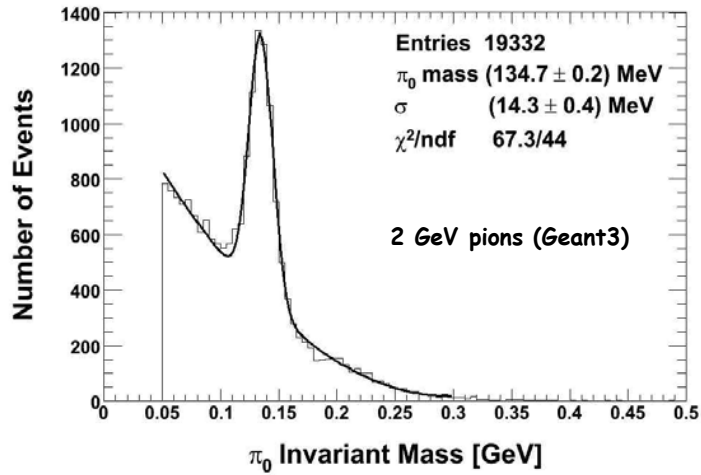

**FIGURE 5.** Invariant mass of simulated $\pi^0$'s using GEANT-3. The pions are generated at random angle with respect to the calorimeter.

Studies using real beam conditions were also carried out at the M11 pion facility at TRIUMF. Beams of pions, muons and electrons with energy ranging from 120 MeV to 300 MeV were aimed at a full-scale 4 m long prototype of a BCAL module[2]. Trigger counters placed between the source of the beam and the detector were used for timing and trigger. Regular Burle PMT's 8575 photo-sensors were used to collect light from the prototype. Figure 6 shows some preliminary results of an analysis to estimate the attenuation length of the fibers and the time resolution achieved in the tests[3]. While the attenuation length of 309±13 cm is consistent with the manufacturer specifications, it is expected that the time resolution of 320±2 ps will significantly improve with the use of the final version of the photo-sensors (possibly SiPM's), fast green scintillator fibers and improved readout electronics.

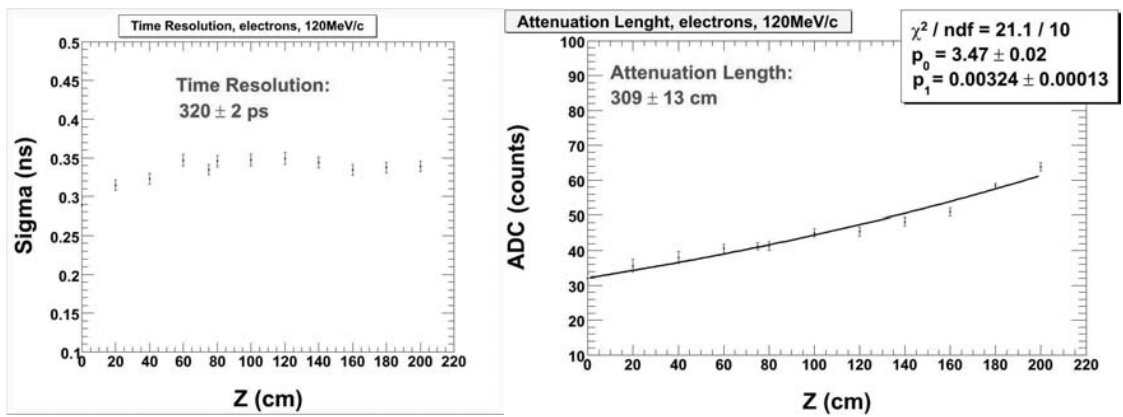

**FIGURE 6.** Preliminary measurements of the attenuation length and time resolution with data from tests using 120 MeV electrons from the M11 beam facility at TRIUMF.

---

[2] Pol.Hi.Tech. double-clad blue scintillator fibers were used for these tests.
[3] These measurements are part of the Master thesis being prepared by Mrs. Gergana Koleva at the University of Regina.

As discussed before, SiPM's have being considered as the best candidates to fulfill the requirement of operation in a ~2 T magnetic field, and deliver high energy and time resolution. Currently, only small devices with active area of 1x1 mm$^2$ are available in the market. Given the proposed BCAL readout scheme, as discussed in the previous sections, these small devices cannot be used to instrument the calorimeter. However, the GlueX collaboration has entered into an agreement with the SensL Company[4] for the development of customized large area photo-sensors consisting in arrays of SiPM's for the BCAL-GlueX detector. Each photo-sensor is a matrix of 3x3 mm$^2$ SiPM's. A study on the dependence of the minimum number of detectable photoelectrons as a function of the photo-sensor area and dark rate is shown in Figure 8 based on simulations of the SiPM parameters. This study was carried out by the SensL group using two values for the SiPM dark rate at one photoelectron level: 3 MHz and 5 MHz. The total photo-sensor dark rate is the sum of the dark rate of each SiPM in the array. The SiPM's were not optimized to low optical cross-talk effect and no cooling system was considered into the simulation. A minimum number of 170 photoelectrons is expected to reach the lower 20 MeV energy threshold level with the calorimeter. As can be observed, 4x4 arrays of 3x3 mm$^2$ SiPM's, equivalent to photo-sensors with 12x12 mm$^2$ active area, can deliver a minimum of 160 photoelectrons at 80 MHz dark rate level (5 MHz/SiPM). These parameters will be significantly improved with the optimization of the SiPM's for low optical cross-talk effect and the use of a cooling system which will further decrease the dark rate to even lower levels with consequent increase in the minimal number of detectable photoelectrons.

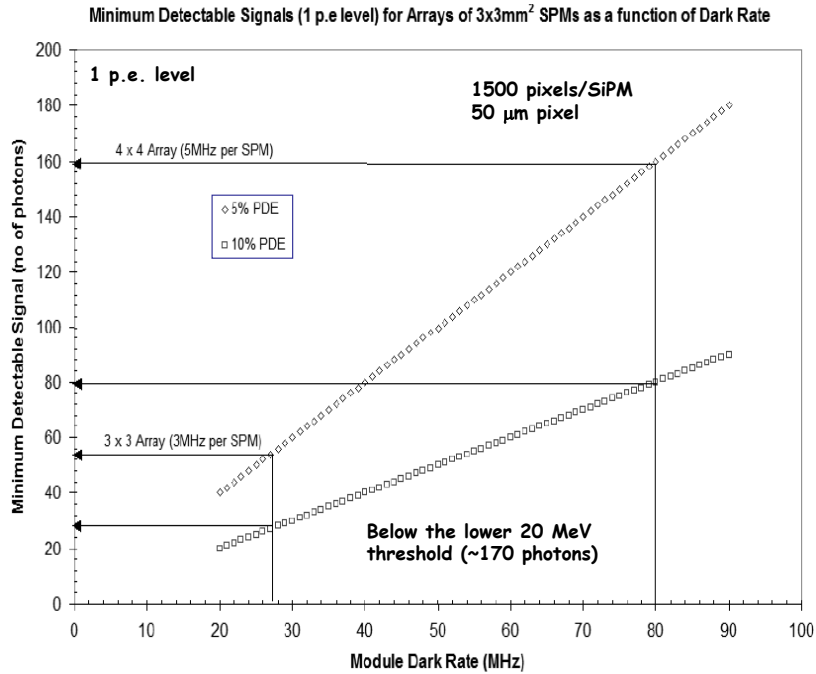

**FIGURE 7.** Minimum number of detectable photoelectrons as a function of the photo-sensor array area and total dark rate.

---

[4] The Low Light Sensing Company (SensL), http://www.sensl.com/.

In order to collect light from individual calorimeter readout areas of 20x20 mm$^2$ (inner layers), each photo-sensor will be coupled to Winston cones with maximal geometrical reduction factor of 4 times.

## Conclusions

This contribution summarizes some of several studies for the development of the electromagnetic calorimeter for the GlueX detector. These studies have shown that the project of a large acceptance non-compensating SciFi-Pb calorimeter using silicon-photomultiplier based photo-sensors for readout is feasible. High energy and time resolution can be achieved with the proposed geometrical configuration, leading to electromagnetic shower reconstruction to high accuracy as required for the success of the GlueX physics programs.

## ACKNOWLEDGMENTS

The author would like to thank the GlueX Collaboration for the material presented in this contribution. Particular thanks goes to Mrs. Gergana Koleva for providing some of her master thesis results to be included in this paper, to Dr. Rafael Hakobyan and Mr. Blake Leverington for the Monte Carlo simulations, to Dr. Zisis Papandreou and Dr. George Lolos for several calculations and discussions and finally to SensL Company for the information concerning the development of the photo-sensors. This research is supported by the National Sciences and Engineering Research Council of Canada and U.S. Department of Energy.

## REFERENCES

1. C.J. Morningstar and M. Peardon, *Phys. Rev. D* **60**, 034509 (1999).
2. G.S. Bali, K. Schilling and C. Schlichter, *Phys. Rev. D* **52**, 5165 (1995) .
   T. Burns, F.E. Close, *Phys. Rev. D* **74**, 034003 (2006).
3. A.R. Dzierba, "QCD Confinement and the Hall D Project at Jefferson Lab" in *e$^+$e$^-$ Physics at Intermediate Energy Worshop-2001*, eConf C010430 T04, 2001. Also in e-Print Archive hep-ex/0106010.
    GlueX Collaboration Homepage: http://gluex.org.
4. V.D. Kovalchuck, G.J. Lolos, Z. Papandreou, K. Wolbaum, *Anucl. Instrum. Meth.* **A538**, 408-415 (2005).
5. E. Garutti, M. Groll, A. Karakash and S. Reiche, ALC-DET-2004-025, (2004).
6. P. Buzhan *et al*, *ICFA Instrum. Bull.* **23**, 28-41 (2001).